\RequirePackage{amsmath}

\documentclass[12pt]{iopart}
\pdfoutput=1
\usepackage{graphicx}
\usepackage{amsmath,amsfonts,amssymb}

\usepackage{xcolor}
\usepackage{tikz}
\usetikzlibrary{calc,arrows} 
\colorlet{mylinkcolor}{blue!66!black!80}

\usepackage[colorlinks=true,linkcolor=mylinkcolor,citecolor=mylinkcolor,filecolor=cyan,urlcolor=mylinkcolor,breaklinks=true]{hyperref}

\usepackage[utf8]{inputenc}

\usepackage{cite}

\newcommand{\psiL}{\psi^\mathrm{L}}
\newcommand{\psiR}{\psi^\mathrm{R}}
\newcommand{\dd}{{\rm d}}

\begin{document}
\title[Duality Between Relaxation and First Passage in Reversible Markov Dynamics]{Duality Between Relaxation and First Passage in Reversible Markov Dynamics: Rugged Energy Landscapes Disentangled}
 \author{David Hartich and Alja\v{z} Godec}
\address{Mathematical Biophysics Group, Max-Planck-Institute for
  Biophysical Chemistry, G\"{o}ttingen 37077, Germany}
\eads{\href{mailto:david.hartich@mpibpc.mpg.de}{david.hartich@mpibpc.mpg.de},\href{mailto:agodec@mpibpc.mpg.de}{agodec@mpibpc.mpg.de}}

\begin{abstract}
 Relaxation and first passage processes are the pillars of kinetics in condensed matter, polymeric and single-molecule systems. Yet, an explicit connection between relaxation and first passage time-scales so far remained elusive. Here we prove a duality between them in the form of an interlacing of spectra. In the basic form the duality holds for reversible Markov processes to effectively one-dimensional targets. The exploration of a triple-well potential is analyzed to demonstrate how the duality allows for an intuitive understanding of first passage trajectories in terms of relaxational eigenmodes. More generally, we provide a comprehensive explanation of the full statistics of reactive trajectories in rugged potentials, incl. the so-called `few-encounter limit'. Our results are required for explaining quantitatively the occurrence of diseases triggered by protein misfolding.
\end{abstract}
 
% \maketitle 

\date{\today}

\section{Introduction}

Relaxation dynamics are a paradigm for describing complex
dynamical phenomena spanning condensed matter \cite{datt87}, polymeric
\cite{doi86}, granular \cite{edwa94} and single-molecule
systems \cite{noe11,chen07}, and even cellular regulatory
networks \cite{walc12}. Relaxation
concepts underlie most spectroscopic methods \cite{datt87}. Moreover, our understanding of
metastability is built entirely on the properties of relaxation
spectra \cite{lang69,bovi15,biro01,tana03,tana04}. Complementary to relaxation processes     
are the statistics of the first passage time, the time a random process
reaches a prescribed threshold value for the first time. First passage time
statistics in turn
are central to
the kinetics of chemical
reactions
\cite{haen90,redn01,metz14a,kope88,szab80,beni10,ben93,osha95,beni14,guer16,meji11},
signaling in biological cells
\cite{beni14,guer16,meji11,holc14,gode16a,gode15,gode16,gode17,greb16,greb17,greb18,vacc15},
transport in disordered media \cite{avra00}, the foraging behavior of bacteria and animals \cite{berg93,bell90,paly14}, up to the spreading
of diseases \cite{lloy01,hufn04} or stock market dynamics
\cite{mant00}. Further important applications of first passage concepts include the
persistence properties in non-equilibrium systems \cite{bray13,maju01,maju02} and stochastic
thermodynamics \cite{neri17,garr17,rold17,ging17,fuch16,nguy17}. 

Both relaxation and first passage processes are essential for theories
building on a diffusive exploration of (free) energy landscapes $U(x)$
\cite{wale04}, which have proven to be particularly invaluable in explaining
the kinetics of chemical reactions \cite{kram40,schu81}, protein
dynamics \cite{frau91,onuc97} incl. recent single protein folding experiments \cite{neup16}, and the dynamics of supercooled
liquids and glasses \cite{biro01,tana03,tana04,garr02}. Whereas relaxation
processes can be understood intuitively in terms of the
eigenmodes and eigenvalues of the underlying Fokker-Planck or Kramers operators
\cite{biro01,tana03,tana04,bark14}, general, and in particular intuitive results about the
full first passage time statistics are much sparser,
and currently do not reach beyond a crude division between so-called
direct and indirect first-passage trajectories for the simplest smooth
potential landscapes
\cite{beni10,beni14,gode16a,gode16,gode17}. Our
general understanding of first passage phenomena would therefore substantially
benefit from a deeper connection to the corresponding
relaxation process.

Indeed, in the limit of high energy barriers
a well-known link relates $\lambda_1^{-1}$, the longest relaxation time, and the mean
first passage time to surmount the highest barrier in the landscape
\cite{lang69,bovi15,biro01,tana03,tana04,matk81}. However, in spite of the
immense success and universal applicability of this approximate
relationship,
an explicit bridge between first passage and relaxation time-scales has not been
explored further.  

Here we establish such a link rigorously for microscopically reversible Markovian dynamics. We prove that the first-passage
process to an effectively one-dimensional target is in fact
the \emph{dual} to the corresponding relaxation process. The duality takes the form of a
\emph{spectral interlacing} of characteristic time-scales, in which each pair of successive relaxation time-scales
encloses a first-passage time-scale. 
We establish an explicit
relationship between relaxation and first-passage spectra, and
express the full statistics of first passage time exactly in terms of
the relaxation eigensystem. As a case study we consider a diffusive
exploration of a triple-well
potential. Moreover, exploiting the duality we disentangle first passage time
statistics in general rugged energy landscapes. We argue why our results are
important for a quantitative understanding of the occurrence of
diseases related to protein misfolding.

The paper is organized as follows. In Sec.~\ref{sec:main}
we expose the duality between first passage and relaxation processes.
In Sec.~\ref{sec:triple} we determine the first
passage time statistics in a simple triple-well potential and demonstrate
that knowing the full probability density is mandatory
in studies of many-particle first passage problems in the few encounter limit.
In Sec.~\ref{sec:rugged} we determine the first passage time density
for a truly rugged energy landscape generated by a truncated Karhunen-Lo\`eve expansion of a Wiener process.
The large deviation limit of the first passage time distribution,
which is relevant for single-molecule first passage problems, is
presented in Sec.~\ref{sec:mu1}. We conclude in Sec.~\ref{sec:conclusion}.
A proof of the duality between first passage and relaxation is
relegated to \ref{Appendix1}.

\section{First passage time density from the relaxation spectrum}
\label{sec:main}
We consider reversible Markovian dynamics in
continuous time governed by a Fokker-Planck
operator $\mathbf{L}=\partial_x D(x)[\beta U'(x)+\partial_x]$, where $x$ is the position,  $U(x)$ a potential with $U'(x)\equiv \partial_x U(x)$, and $D(x)$ the diffusion landscape.
We assume $\beta$ to be the inverse temperature, such that  according to the fluctuation-dissipation theorem $\beta D(x)$  is the inverse  friction coefficient.
For any initial condition $x_0$ the dynamics governed by the Fokker-Planck operator $\mathbf{L}$ relaxes to the Boltzmann distribution $P_{\mathrm{eq}}(x)=\mathrm{e}^{-\beta
  U(x)}/\int\mathrm{e}^{-\beta U(x)}\dd x$. Adopting the bra-ket notation we expand $\mathbf{L}$ in a
complete bi-orthogonal set of left and right eigenstates,
$\mathbf{L}=-\sum_{k}\lambda_k|\psiR_k\rangle \langle\psiL_k|$, $\lambda_k$ denoting the eigenvalues and  $\langle\psiL_k|\psiR_l\rangle=\delta_{kl}$ with $|\psiL_k\rangle\equiv\mathrm{e}^{\beta U(x)}|\psiR_k\rangle$. The propagator encoding the probability to be at $x$ at a time $t$ after
starting from $x_0$ at $t_0=0$, is defined as
\begin{equation}
  P(x,t|x_0)\equiv
  \langle x|
\mathrm{e}^{\mathbf{L}t}|x_0\rangle=\sum_{k}\langle x |\psiR_k\rangle\langle\psiL_k|x_0\rangle\mathrm{e}^{-\lambda_kt}.
\label{propagator}
 \end{equation} 
Since we assumed temporally homogeneous dynamics, we can
define the first passage time probability density from some $x_0$ to a
target at $a$, $\wp_a(t|x_0)$, by the renewal theorem \cite{sieg51}
\begin{equation}
  P(x,t|x_0)=\int_0^t\wp_a(\tau|x_0)P(x,t-\tau|a)d\tau,
\label{renewal}
\end{equation}
where either $x_0<a\le x$, or symmetrically $x_0>a\ge x$. Eq.~(\ref{renewal})
follows from a direct enumeration of paths between $x_0$ and $x$,
which by construction must pass through $a$. 
Laplace transforming Eq.~(\ref{renewal}) we obtain
$\tilde{\wp}_a(s|x_0)=\tilde{P}(x,s|x_0)/\tilde{P}(x,s|a)$, which is
the starting point of our analysis. Using Eq.~(\ref{propagator}),
which after Laplace transform reads
$\tilde P(x,s|x_0)=\sum_{k}(s+\lambda_k)^{-1}\langle x |\psiR_k\rangle\langle\psiL_k|x_0\rangle$,
yields
\begin{equation}
\tilde{\wp}_a(s|x_0)=\frac{\sum_{k}(s+\lambda_k)^{-1}\langle x |\psiR_k\rangle\langle\psiL_k|x_0\rangle}{\sum_{k}(s+\lambda_k)^{-1}\langle x |\psiR_k\rangle\langle\psiL_k|a\rangle}.
\label{fpt}
 \end{equation}
The Laplace transform of the first passage time density
$\tilde{\wp}_a(s|x_0)$ is a meromorphic function having simple poles $-\mu_k$
on the negative real axis \cite{keil64}. Moreover, the poles have no accumulation point in the left half plane
($\mathrm{Re}(s)<0$). Similarly, $\tilde{P}(y,s|x)$ is meromorphic
with simple poles $-\lambda_k$ arranged along the non-positive real
axis. In particular, $\lambda_0=0$ and $\langle x |\psiR_0\rangle\langle\psiL_0|x_0\rangle=P_{\mathrm{eq}}(x)$. 

The Laplace transforms of the propagators $\tilde{P}(x,s|x_0)$ and $\tilde{P}(x,s|a)$ have
coinciding poles, while the poles of $\tilde{\wp}_a(s|x_0)$
are those zeroes of $\tilde{P}(x,s|a)$, which are different from the
zeroes of $\tilde{P}(x,s|x_0)$. Generally, $\tilde{P}(x,s|x_0)$ and $\tilde{P}(x,s|a)$ have
infinitely many coinciding zeroes alongside the distinct ones (see
proof in \cite{hart18a_arxiv}), because the region beyond $a$
cannot affect the first passage time from $x_0$, whereas it must affect the
relaxation. However, all common zeroes result in a
vanishing residue.

One can prove that setting
$x=a$ in Eqs.~(\ref{renewal}) and (\ref{fpt}) guarantees that
\emph{all relevant} eigenvalues (i.e., those satisfying $\langle a|\psiR_k\rangle\neq0$)
of the relaxation and first passage processes interlace
\begin{equation}
  \lambda_{k-1} < \mu_k <\lambda_{k}, \quad \forall k\ge1,
  \label{interlacing}
\end{equation}
which is due to the fact that $\langle a |\psiR_k\rangle\langle\psiL_k|a\rangle>0$ (see also \cite{hart18a_arxiv}).
Based on the interlacing in Eq.~(\ref{interlacing}) we are now in the position to
determine the entire first passage time statistics from the relaxation eigenspectrum, $\{|\psiR_k\rangle,\langle\psiL_k| ,\lambda_k\}$. The calculation becomes rather involved
and is sketched in \ref{Appendix1} (see also \cite{hart18a_arxiv}), here
we simply state the result. Introducing
$\bar{\mu}_k=(\lambda_k+\lambda_{k-1})/2$ the corresponding first passage eigenvalues $\mu_k$
are given exactly in the form of a convergent Newton's series
\begin{equation}
\mu_k=\bar{\mu}_k+\sum_{n=1}^{\infty}f_0(k)^nf_1(k)^{1-2n}\frac{\mathrm{det}\boldsymbol{\mathcal{A}}_n(k)}{(n-1)!},
\label{eigenv}  
\end{equation}
with the almost triangular $(n-1)\times(n-1)$  matrix $\boldsymbol{\mathcal{A}}_n(k)$
with elements
\begin{equation}
\label{element}
\mathcal{A}^{i,j}_n(k)=\frac{f_{(i-j+2)}(k)\Theta(i-j+1)}{(i-j+2)!}\Big[n(i-j+1)\Theta(j-2)
+i\Theta(1-j)+j-1\Big],
\end{equation}
where $\Theta(l)$ denotes the discrete Heaviside step function ($\Theta(l)=1$ if $l\ge0$), and
symbolically we set $\mathrm{det}\boldsymbol{\mathcal{A}}_1\equiv 1$. 
Setting $k^{\ast}=k$ if $\tilde{P}(a,-\bar{\mu}_k|a)<0$ and
$k^{\ast}=k-1$ otherwise, $f_n(k)$ in Eqs.~(\ref{eigenv}-\ref{element}) are defined by
\begin{equation}
\label{coefs}  
\begin{aligned}
 f_0(k)&=\langle a |\psiR_{k^{\ast}}\rangle\langle\psiL_{k^{\ast}}|a\rangle+\sum_{l\ne k^{\ast}}{\langle a |\psiR_l\rangle\langle\psiL_l|a\rangle\frac{(\bar{\mu}_k-\lambda_{k^{\ast}})}{(\bar{\mu}_k-\lambda_l)}},\\
f_{n\ge 1}(k)&=n!\sum_{l\ne k^{\ast}}{\langle a |\psiR_l\rangle\langle\psiL_l|a\rangle \frac{(\lambda_l-\lambda_{k^{\ast}})}{(\bar{\mu}_k-\lambda_l)^{n+1}}}.
\end{aligned}
\end{equation}
Using Eq.~(\ref{coefs}) the determinant $\det \boldsymbol{\mathcal{A}}_n(k)$ of the almost triangular matrix with elements \eqref{element} is fully characterized, which in turn
determines the first passage eigenvalue Eq.~\eqref{eigenv}.
 It is now easy to obtain $\wp_a(t|x_0)$ by inverting the Laplace
transform in Eq.~(\ref{fpt}) using Cauchy's residue theorem yielding
\cite{hart18a_arxiv}
\begin{equation}
\wp_a(t|x_0)=\sum_{k\ge 1}w_k(x_0)\mu_k\mathrm{e}^{-\mu_kt},  
\label{first}
\end{equation}
where the spectral weights as a function of  the initial
condition $x_0$, $w_k(x_0)$, are given by
\begin{equation}
w_k(x_0)=\frac{\sum_{l}(1-\lambda_l/\mu_k)^{-1}\langle a |\psiR_l\rangle\langle\psiL_l|x_0\rangle}{\sum_{l}(1-\lambda_l/\mu_k)^{-2}\langle a |\psiR_l\rangle\langle\psiL_l|a\rangle},
\label{weight}
\end{equation}
where we only sum over relevant $\lambda_l$, and the weights are normalized $\sum_{k\ge 1}w_k(x_0)=1$. Moments of the first passage time follow immediately,
\begin{equation}
 \langle t^n_a(x_0)\rangle=n!\sum_{k\ge 1}w_{k}(x_0)\mu_k^{-n}
 \label{eq:moments}
\end{equation}
 We note that while the first weight is necessarily positive, $w_1(x_0)>0$, the other weights $w_k(x_0)$ can be negative for $k>1$. In the following section
 we will show how this duality between first passage and relaxation can
 be used to determine and understand more deeply the full first passage time distribution.
 
 For the sake of completeness we briefly comment on the ``reverse'' direction of the duality. The starting point is the renewal theorem \eqref{renewal} for $x=a$
 in Laplace space, which reads $\tilde{\wp}_x(s|x_0)=\tilde P(x,s|x_0)/\tilde P(x,s|x)$.
 Using standard Green's function theory \cite{hart18a_arxiv} it can be shown after some straightforward but tedious algebra
 that 
 \begin{equation}
  \tilde P(x,s|x_0)=\sigma_{\pm}\frac{\e^{\beta U(x_0)-\beta U(x)}\tilde \wp_{x_0}(s|x)}{D\frac{\partial}{\partial x_0}\ln[\tilde\wp_{x_0}(s|x)\tilde\wp_{x}(s|x_0)]}
  \label{dualityInv}
 \end{equation}
holds, where $\sigma_{\pm}=-1$ if $x_0<x$ and $\sigma_{\pm}=+1$ if $x_0>x$.
For the remainder of the paper we will focus solely on the explicit forward duality, since it allows us to efficiently determine the full first passage time density.

\section{Triple-well potential}
\label{sec:triple}
\subsection{First passage time density}

\begin{figure}
\includegraphics[width=\textwidth]{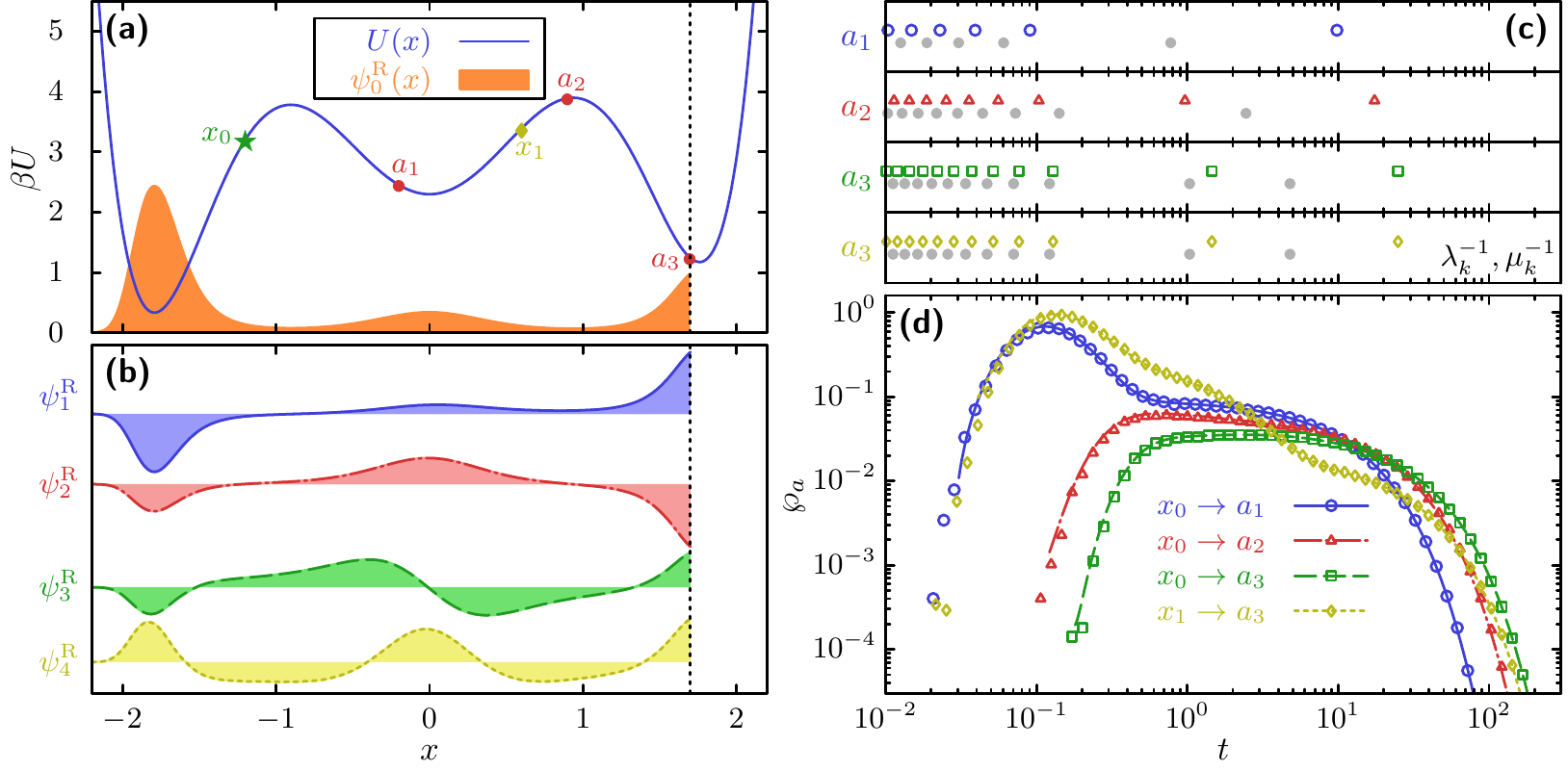}
\caption{(a) Triple well potential $\beta U(x)=(x^6 -6x^4 + 0.15x^3 +
  8x^2)/2$ (line) and corresponding invariant measure
  $P_{\mathrm{eq}}(x)\propto\psiR_0(x)$ (shaded). Highlighted are the initial
  conditions $x_0=-1.2$ and $x_1=0.6$ alongside three target positions
  $a_1=-0.2$, $a_2=0.9$ and $a_3=1.7$; (b)
  Right eigenvectors $\psiR_k(x)\equiv\langle x |\psiR_k\rangle$ for the four slowest
eigenmodes of the relaxation process with a reflecting boundary at
$a_3$; (c) Spectra of characteristic time scales for relaxation
($\lambda_k^{-1}$, filled gray circles), and first
passage ($\mu_k^{-1}$, open color symbols) processes, with the boundaries at $a_1$ to $a_3$, respectively; we used relflecting boundaries for the determination of $\lambda_k^{-1}$ and absorbing boundaries for $\mu_k^{-1}$; (d) First passage time probability
densities -- the lines correspond to Eq.~(\ref{first}) and the symbols to
Brownian dynamics simulations each of an ensemble of $2\times 10^6$ trajectories with an integration step $\Delta t=10^{-5}$.}
\label{fg1}
 \end{figure}

 As a case
 study we analyze the first passage time statistics in a triple
 well potential (see Fig.~\ref{fg1}a). Understanding the diffusive exploration of multi-well
 potentials is important from a biophysical perspective, as it
 underlies e.g. the folding \cite{neup16,bryn89}, misfolding
 \cite{yu15,dee16}, conformational dynamics \cite{noe11,chen07} and
 aggregation of proteins and peptides \cite{dobs03,zhen13} as well as (bio)chemical
 reactions \cite{kram40,schu81}.
 
 We computed the first 40 left and right
 relaxation eigenvectors, $\langle x|\psiR_k\rangle$ and $\langle
 \psiL_k|a\rangle$, and eigenvalues $\lambda_k$ numerically using a reflecting boundary condition at the target $a$.
 The four lowest $\langle x|\psiR_k\rangle$  of the relaxation process are depicted in Fig.~\ref{fg1}b. From
 $\{\lambda_k,|\psiR_k\rangle,\langle \psiL_k|\}$ we calculate the first 30 $\mu_k$ and
 $w_k(x_0)$ using the duality, i.e. Eqs.~(\ref{eigenv}) and
 (\ref{weight}). The spectrum of first passage eigenvalues is depicted in
 Fig.~\ref{fg1}c, with the corresponding first passage time probability densities shown in
 Fig.~\ref{fg1}d (lines) and compared to the result of Brownian dynamics
 simulations (symbols). We find an excellent agreement between theory
 and simulations. Note that the deviations of the theoretical results
 from simulations observed on extremely short timescales are a direct
 consequence of truncating the sums in Eqs.~(\ref{propagator}) and
 (\ref{first}) (i.e., we considered 40 eigenvalues in the relaxation spectrum and 30 first passage eigenvalues).
 
 We now link metastability to the
 first passage time behavior. A potential $U(x)$ has metastable states if the minima
 are separated by high barriers $>k_\mathrm{B}T$. The probability mass in the
 ground state $P_{\mathrm{eq}}(x)$ is concentrated around these
 minima. The barriers give rise to a separation of time-scales between 
inter-well (see, e.g., $\psiR_1,\psiR_2$ in Fig.~\ref{fg1}b) and intra-well dynamics (see $\psiR_{k>2}$),
 and thus create gaps in the
 relaxation spectrum \cite{bovi15,biro01,tana03,tana04}. As a
 result we observe in Fig.~\ref{fg1}c (see filled gray circles) two gaps $0=\lambda_0\ll\lambda_1\ll\lambda_2<\lambda_3$ when
 the reflecting boundary is at $a_1$, corresponding to the crossing of
 a single barrier. Conversely, three gaps, $0\ll\lambda_1\ll\lambda_2\ll\lambda_3<\lambda_4$, appear when the
 reflecting boundary is at $a_3$, corresponding to the global
 relaxation to $P_{\mathrm{eq}}(x)$, to direct transitions between the leftmost and
 right-most wells, and to
 the transition to the central well from both
 sides, respectively (see Fig.~\ref{fg1}b). These gaps are independent of $x_0$.

\begin{figure}
 \centering
\includegraphics[width=\textwidth]{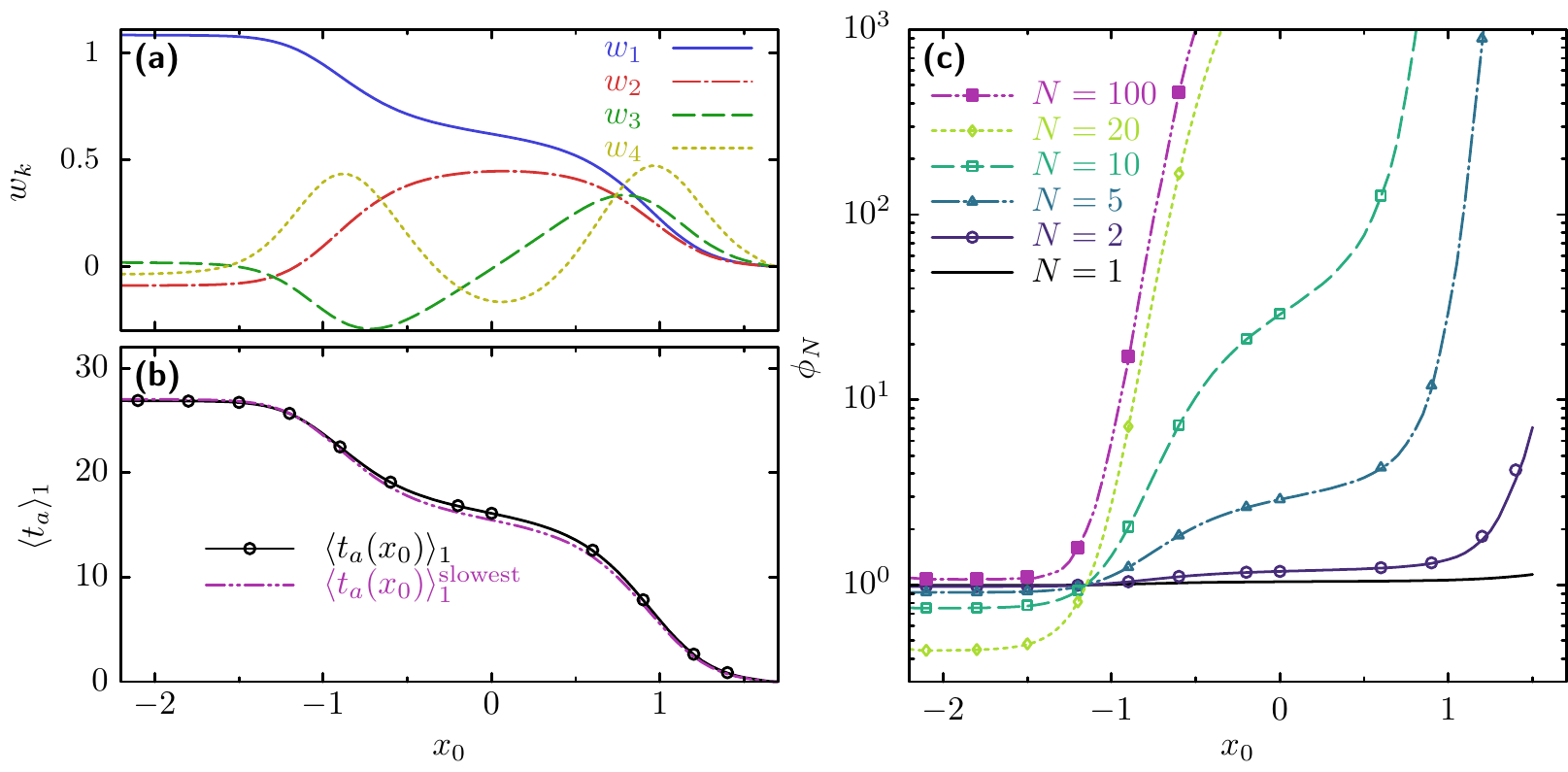}
\caption{First passage in  triple-well potential from Fig.~\ref{fg1} to fixed target position $a_3=1.3$.
(a)~Lowest four weights $w_k(x_0)$ determined from Eq.~(\ref{weight}) for the
 first passage process from $x_0$ to fixed target $a=a_3$ in the triple-well potential depicted in  Fig.~\ref{fg1}a  as a function of the starting point $x_0$.
(b)~Mean first passage time (solid black line) corroborated by Brownian simulations (symbols) versus slowest time-scale approximation $\langle t_a \rangle_1^{\rm slowest}=w_1(x_0)/\mu_1$ from Eq.~(\ref{tslowest}).  (c)~$\phi_N=\langle t_a \rangle_N/\langle t_a \rangle_N^{\rm slowest}$ comparing the
  mean first passage time
  of $N$ particles, $\langle t_a (x_0)\rangle_N$, with
  the one-scale approximation  $\langle t_a (x_0)\rangle_N^{\rm slowest}=w_1(x_0)^N/(N\mu_1)$ (see Eq.~(\ref{tslowest})) for
  various $N$. Lines denote the theory and symbols the
  simulation results.}
\label{fg_w}
\end{figure}

 Due to the interlacing (Eq.~(\ref{interlacing})),
 and because $\mu_1\ne 0$, the $N$ gaps in the relaxation spectrum reflecting all the metastable basins translate to $N-1$
 gaps in the first passage spectrum due to the $N-1$ barriers. The first passage spectrum is shifted to shorter
 times, since contrary to relaxation, \emph{all trajectories} must surmount the
 barriers. The spectral weights $w_k$ depend on the initial
 position, gauging the contribution of each relaxation mode with respect to the
 given first passage time-scale $\mu_k^{-1}$ (see Eq.~(\ref{weight})). The four
 lowest $w_k$ for the first passage process $x_0\to a_3$ are shown in
 Fig.~\ref{fg_w}a.     

In view of \cite{gode16,gode17} (see also \cite{beni14}) we now separate
all first passage trajectories into two classes --- the so-called
`globally indirect' and the rest.
The class of `globally indirect' trajectories
includes those exploring the entire accessible phase space prior to
absorption. These trajectories therefore arrive on the slowest
time-scale $\mu_1^{-1}$ and their associated
weight $w_1$ is approximately the fraction of all first passage trajectories that
reach quasi-equilibrium before hitting the target. Correspondingly,
$w_1$ --  the weight of globally indirect trajectories decreases as
the starting position $x_0$ approaches the target at $a$ (see e.g. blue
solid line in Fig.~\ref{fg_w}a for $a=1.7$). In other words, 
the closer $x_0$ is to the target the more unlikelier are globally
indirect trajectories.

Pushing this picture even further we can also identify in Fig.~\ref{fg1}d ($x_1\to a_3$)
a second pronounced time-scale $\mu_2^{-1}$ with weight $w_2$,
reflecting what we may call `locally indirect' trajectories -- those that first equilibrate locally
within the central well but cross the second barrier
without returning to the left, deepest well.
Comparing the second weight $w_2$ from Fig.~\ref{fg_w}a (see dash-dotted red line)
and the potential landscape Fig.~\ref{fg1}a we find that `locally indirect'
trajectories are most pronounced in the sense of the largest value of $w_2$ if the starting position is within the central well.
The locally indirect trajectories
account for local equilibration prior to absorption and become
relevant as soon as the potential landscapes has more than one deep free energy
basin, such as for example the one depicted in Fig.~\ref{fg1}. Our work therefore
extends the present understanding of first passage processes
\cite{beni14,gode16,gode17} by explicitly identifying
locally indirect trajectories -- those equilibrating only locally prior to absorption.

For $x_0$ within
the central well the fraction of globally indirect trajectories
decreases, and locally indirect trajectories become likelier, i.e.
$w_2$ increases. Concurrently, higher spectral weights also grow,
rendering direct trajectories more likely. As a result, an additional
time-scale appears, giving rise to a second `bump' in $\wp_a(t)$ (see Fig.~\ref{fg1}d, blue
lines). This reasoning extends to arbitrary landscapes; $w_k,\mu_k$ reflect a hierarchy of time-scales, on which trajectories equilibrate locally in
the sequence of all intervals between consecutive basins and $a$,
before hitting $a$. The highest modes encode direct trajectories.

\subsection{Few encounter kinetics require the full first passage time distribution}

The full first passage time statistics are crucial for kinetics in the few-encounter
limit, when only the first of many
particles needs to find the target  \cite{gode16,gode17}.
We highlight this
on hand of first passage time statistics in a non-interacting $N$-particle
system. The $N$-particle survival probability --- the probability that
none of the $N$ particles starting from $x_0$ has reached the target
until time $t$ ---
is simply given by
\begin{equation}
\label{survival}
 \mathcal{P}_a(t|x_0)^N\equiv\left[\int_t^{\infty}\wp_a(\tau|x_0)d\tau\right]^N=\left[\sum_{k>0}w_k(x_0)\e^{-\mu_k t}\right]^N,
\end{equation}
where we have inserted Eq.~(\ref{first}). We note that if the initial conditions where not identical with $x_i\neq x_0$ for all $i=1,\ldots, N$ one would replace
the survival probability
$\mathcal{P}_a(t|x_0)^N$ by the product $\prod_{i=1}^N \mathcal{P}_a(t|x_i)$.
For convenience, we will restrict our discussion to the scenario in which all particles start from the same position.
Using the survival probability \eqref{survival} the $N$-particle
first passage time density follows directly from the single particle case
\begin{equation}
 \wp_a^{(N)}(t|x_0)\equiv-\frac{\partial}{\partial t} \mathcal{P}_a(t|x_0)^N= N\wp_a(t|x_0) \mathcal{P}_a(t|x_0)^{N-1},
 \label{Ndensity}
\end{equation}
which is the probability density that one of $N$ particles
reaches the target $a$ at time $t$ under the condition that none
of the remaining $N-1$ particles has arrived before. Obviously,
$N$-particles will find the target on average in a shorter time than a single particle. More precisely, the mean
first passage time in the many particle setting reads
\begin{equation}
 \langle t_a(x_0)\rangle_N\equiv\int_0^{\infty}t\wp_a^{(N)}(t|x_0)\dd t=\int_0^{\infty}\mathcal{P}_a(t|x_0)^N\dd t,
\end{equation}
where we have inserted Eq.~(\ref{Ndensity}) and performed an integration by parts in the last step.

Let us now focus on the mean first passage time and start with a
single particle exploration $(N=1)$ in which case the mean according
to Eq.~\eqref{eq:moments} is simply given by $ \langle
t_a(x_0)\rangle_1=\sum_{k>0}w_k(x_0)/\mu_k$. If there are free energy
barriers between the initial position of the particle and the target,
which lead the emergence of a local equilibrium before reaching $a$, i.e., $\mu_2\gg\mu_1$, we expect the mean first passage time to be 
well  approximated by the slowest timescale $ \langle t_a(x_0)\rangle_1\approx w_1(x_0)/\mu_1\equiv  \langle t_a(x_0)\rangle_1^\text{slowest}$.
The dominance of the single slowest time-scale is fully corroborated
by simulations as depicted in Fig.~\ref{fg_w}b (compare solid black line and dash-dotted magenta line).
Notably, the approximation $\langle t_a(x_0)\rangle_1\approx w_1(x_0)/\mu_1$ $(\ll w_2(x_0)/\mu_2)$
can be accurate even if the barriers are not located between $x_0$ and $a$
as can be seen in Fig.~\ref{fg_w} for  $x_0\gtrsim 1$ 
(see Fig.~\ref{fg1}a for potential landscape). This can be explained
intuitively by the fact that $w_{1}(x_0)$ is approximately the splitting probability
that the particle will reach the deepest potential well at
$x^\dagger\approx-1.8$ before hitting the target $a=a_3$,
multiplied by the average time to leave the deepest basin in the potential, which is $1/\mu_1$ \cite{gode16}.

In the $N$ particle setting the ``slowest'' first passage rate is simply $\mu_1^{(N)}=N\mu_1$,
such that the long-time limit of the first passage time density is given by
$\wp^{(N)}_a(t|x_0)\simeq w_{1}^{(N)}(x_0)\mu_1^{(N)}\e^{-\mu_1^{(N)}t}$
with weight $w_{1}^{(N)}(x_0)=w_{1}(x_0)^N$. 
Utilizing only long-time asymptotics in the $N$-particle
system gives
\begin{equation}
 \langle t_a(x_0)\rangle_N^\text{slowest}=\frac{w_1(x_0)^N}{N\mu_1}.
 \label{tslowest}
\end{equation}

Comparing the exact $\langle t (x_0) \rangle_N$
with this approximation in terms of  $\phi_N=\langle t (x_0) \rangle_N/\langle t (x_0) \rangle_N^{\rm slowest}$
(see Fig.~\ref{fg_w}c) reveals, however, that the long-time approximation
can be orders
of magnitude off, despite its accuracy in the single particle setting. In particular, it underestimates $\langle t (x_0) \rangle_N$ for
distant $x_0$ up to approximately the point $x_0\approx-1.1$ (see curves with $1\le N\le20$), where $w_2$  changes sign, where from it
overestimates  $\langle t (x_0) \rangle_N$. Increasing $N$ further beyond $N>20$
shift the first passage towards shorter time scales, rendering higher modes corresponding to $w_{k\ge 3}$ more relevant, which finally yields a systematically longer mean first passage time than expected from the single times-scale estimate (see magenta line with filled rectangles, $N=100$, in Fig.~\ref{fg_w}c). The large discrepancy
as a result of neglecting direct and locally indirect trajectories
grows further with increasing $N$, and highlights
the importance of understanding the full first passage time
statistics. The lines in Fig.~\ref{fg_w}c are obtained with the
duality relation presented in Sec.~\ref{sec:main} and are fully corroborated by Brownian dynamics simulations (symbols).

\begin{figure}
 \centering
 \includegraphics{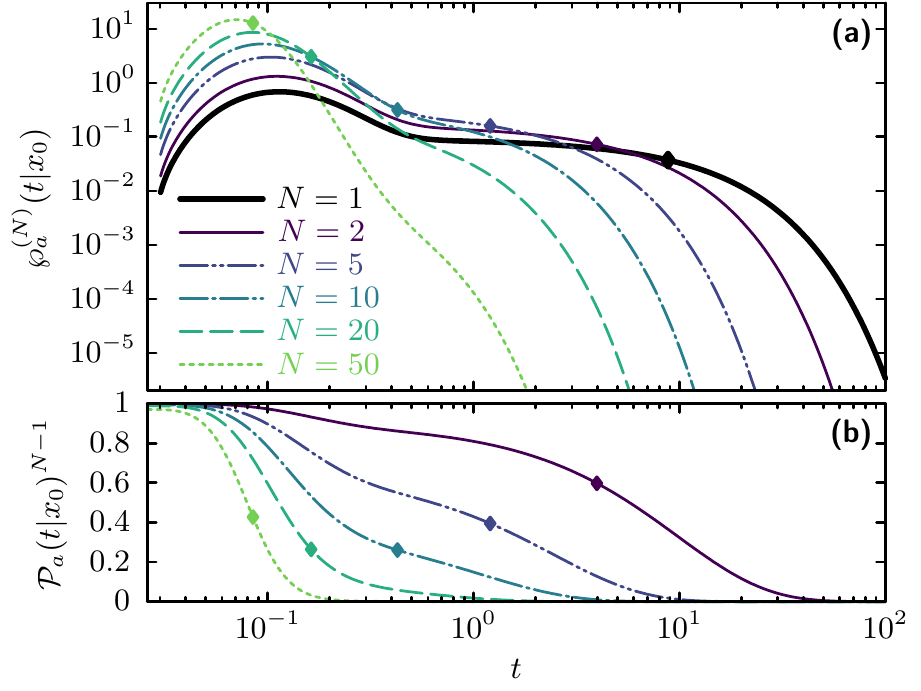}
 \caption{Narrowing of the first passage time density in the
   few-encounter limit. (a)~$N$-particle density for $x_0\to a_1$. The
   thick solid line corresponds to the solid blue line in
   Fig.~\ref{fg1}d. The diamonds depict the respective mean first
   passage times $\langle t_a(x_0)\rangle_N$. (b)~The survival
   probability $P_a(t|x_0)^{N-1}$ that none of the remaining $N-1$ particles has reached the target.
 According to Eq.~(\ref{survival}) each colored curve in (a) is the product of the thick curve $N=1$ and the corresponding survival probability in (b).
 }
 \label{fig_narrowing}
\end{figure}

We now inspect the shape of the distribution upon increasing the
number of particles in Fig.~\ref{fig_narrowing}. We can identify in
Fig.~\ref{fig_narrowing}a two competing effects that eventually lead
to a canonical narrowing of the first passage time distribution for $N\gg 1$. First, according to Eq.~(\ref{Ndensity})
the $N$-particle density is proportional to $\mathcal{P}_a(t|x_0)^{N-1}$, with $\mathcal{P}_a(t|x_0)$ being a strictly
monotonic decaying function that formally satisfies
$\mathcal{P}_a(0|x_0)=1$ and $\mathcal{P}_a(\infty|x_0)=0$. By increasing
the number of particles $N$
the weight
of the survival probability $\mathcal{P}_a(t|x_0)^{N-1}$ is
progressively shifted towards shorter time-scales (see
Fig.~\ref{fig_narrowing}b), i.e., the long-time asymptotics are
shifted towards shorter times for increasing $N$, thereby decreasing the width
of the probability density.
Second,
at short times the single-particle first passage probability
density for generic diffusion process vanishes \cite{kamp93,beij03,redn01}, $\lim_{t\to
  0}\wp_a(t|x_0)=0$. We will refer to this feature as the ``short-time
cutoff'', which according to Eq.~\eqref{Ndensity} prevails for any number of particles $N$. 
Hence the combination of the suppression of long-time asymptotics and
the short-time cutoff
eventually inevitably leads to a narrowing of the first passage time
distribution, irrespective of the details of the underlying dynamics.
Further studies specifically targeting the short time limit of first passage time distributions can be found in \cite{kamp93,beij03}.

In general, the `$N$-particle' first passage problem is essential for describing
nucleation kinetics, since the occurrence of the first
stable nucleus triggers the spontaneous growth of the new phase (see
e.g. \cite{kamp93,beij03}). A particular form thereof is 
the occurrence of
misfolding-triggered protein aggregation resulting in many
diseases \cite{yu15,dee16,dobs03,zhen13}. Namely, in many-protein
systems the free energy minimum does not correspond to a
folded state, but rather to an aggregate of misfolded proteins
\cite{dobs03,zhen13}. Misfolding of a single protein, which indeed occurs by
slow diffusion in a rough energy landscape \cite{yu15,dee16}, seeds aggregation
similar to a nucleation phenomenon. To predict the onset of
aggregation and hence disease from the protein's energy
landscape, an understanding of the full first passage time statistics is required, and
our work provides the foundations to do so. In the following section we briefly
show that our exact theory from Sec.~\ref{sec:main} can also be applied to
systems with truly rugged energy landscapes.

\section{Rugged energy landscapes}
\label{sec:rugged}
In the previous section we have demonstrated that our theory from Sec.~\ref{sec:main}
can readily be used to obtain first passage time densities for
multi-well barrier crossing problems with barrier heights $>k_{\rm
  B}T$. Moreover, we have discussed the few-encounter limit, for which
it is imperative to have access to the full first passage time
distribution, since any attempt to explain many-particle first passage
kinetics by single-particle moments are prone to fail.

\begin{figure}
 \includegraphics{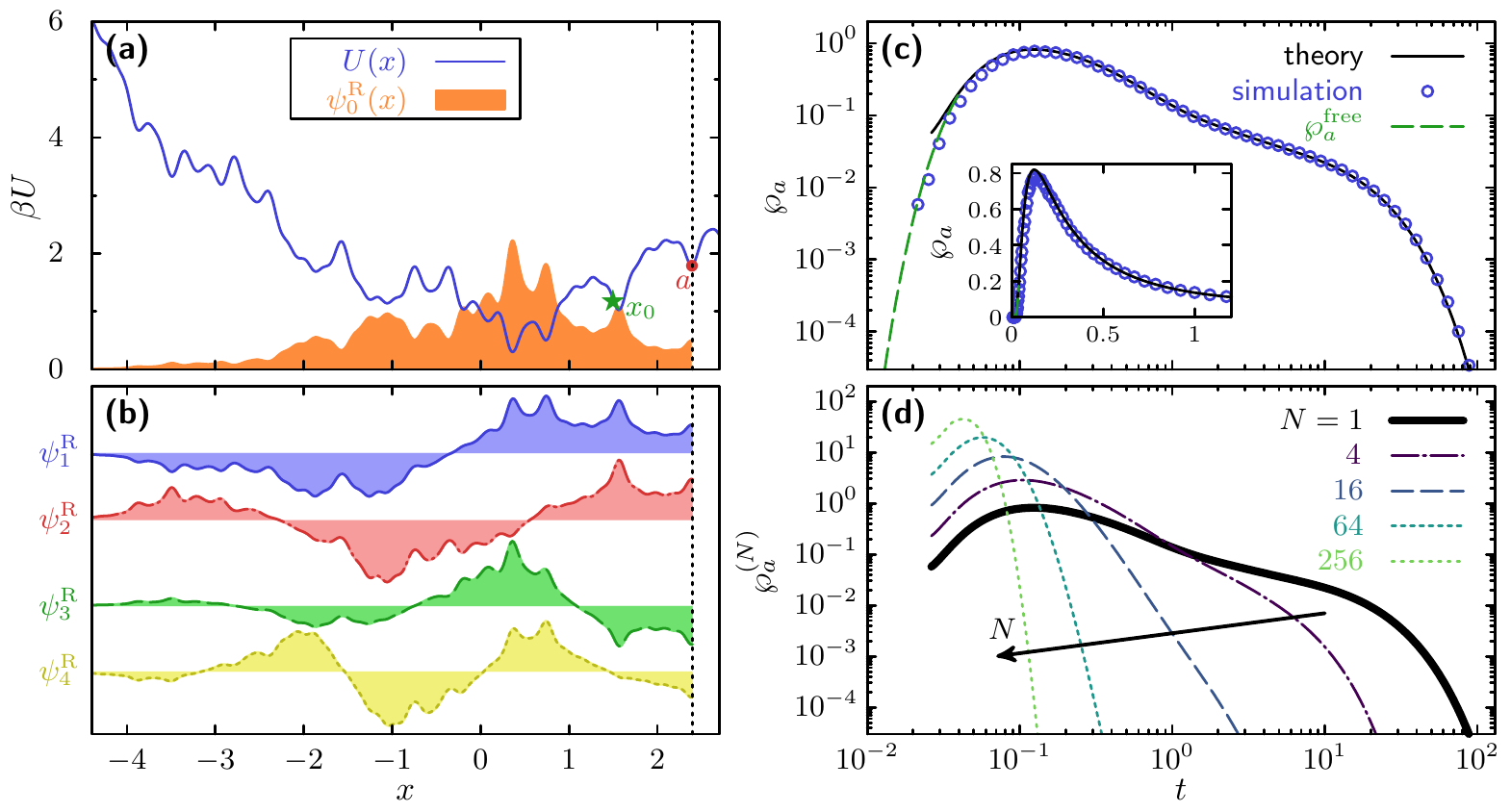}
 \caption{First passage in a rugged energy landscape. (a)~Potential
   from Eq.~(\ref{UKL}) with the corresponding equilibrium measure $\psiR_0(x)\propto P_{\rm eq}(x)$, $K=16$ and
 $(z_1,\ldots,z_{16})=(-0.2,$
 $-0.83,$ $-0.93,$ $1.05,$ $-0.79,$ $0.55,$ $0.61,$ $1.96,$ $-0.88,$ $-1.41,$ $-1.53,$ $-0.31,$ $-0.75,$ $-0.43,$ $-0.46,$ $1.34)$. The first passage from $x_0=1.5$ to $a=2.4$ is considered. (b)~First four excited-state eigenfunctions
   of the relaxation process. (c)~First passage time density
   $\wp_a(t|x_0)$ on a log-log-scale with the corresponding plot on a
   linear scale depicted in the inset. The solid black line is
   determined using the duality Eqs.~(\ref{eigenv}-\ref{weight}) while
   the blue symbols are obtained by simulating $10^6$ trajectories. The
   dashed green line corresponds to the short-time asymptotics of the
   density $\wp_a^\text{free}(t|x_0)=(2\sqrt{\pi t^3})^{-1}|a-x_0|\exp[-(a-x_0)^2/(4t)]$ for a free
   Brownian motion ($U(x)=0$; see, e.g., Ref.~\cite{gode16}). (d)~$N$-particle first passage time density $\wp_a^{(N)}(t|x_0)$.}
 \label{fig_rugged}
\end{figure}

To model a rugged energy landscape containing, in addition to high
barriers, also barriers which are $\lesssim k_{\rm B} T$, we use a parabolic potential
plus a Karhunen-Lo\`eve expansion of a realization of a Brownian motion
\begin{equation}
 U(x)=x^2/4+\sum_{k=1}^Kz_k\frac{\sin[(2k-1)x]}{(2k-1)},
 \label{UKL}
\end{equation}
where we have truncated the potential after $K$ terms and where $z_k$
are Gaussian random numbers. Once $z_k$ are generated we
keep them constant. In Fig.~\ref{fig_rugged}a we depict
the potential generated from Eq.~(\ref{UKL}) with $K=16$.
As before, we determine the eigenvalues $\{\lambda_k\}$
and eigenfunctions $\{\psiR_k(x)\}$ of the relaxation process
with $\psiR_0$ being the equilibrium
Boltzmann density (see left panel of Fig.~\ref{fig_rugged}).
Exploiting the theory from Sec.~\ref{sec:main} we obtain the first passage time density $\wp_a(t|x_0)$ in
Fig.~\ref{fig_rugged}c (see solid black line), which is corroborated by
extensive Brownian dynamics simulations (see blue open circles). The inset of Fig.~\ref{fig_rugged}c depicts the first passage density on a linear scale. In order to indicate the short-time cutoff, which is dominated by diffusive
transport, we also plot the short time asymptotic for free diffusion $(U(x)=0)$. In Fig.~\ref{fig_rugged}d
we depict the corresponding $N$-particle first passage time densities
for the few-encounter limit, which clearly reveal the drastic narrowing of the first
passage time distribution arising from the aforementioned interplay
between the diffusive short-time cutoff and the suppression of the
long-time asymptotics for increasing $N$. This example illustrates
that our theory can readily be applied to arbitrarily rough potential landscapes.

\section{Large deviation limit}
\label{sec:mu1}
\begin{figure}
\centering
\includegraphics{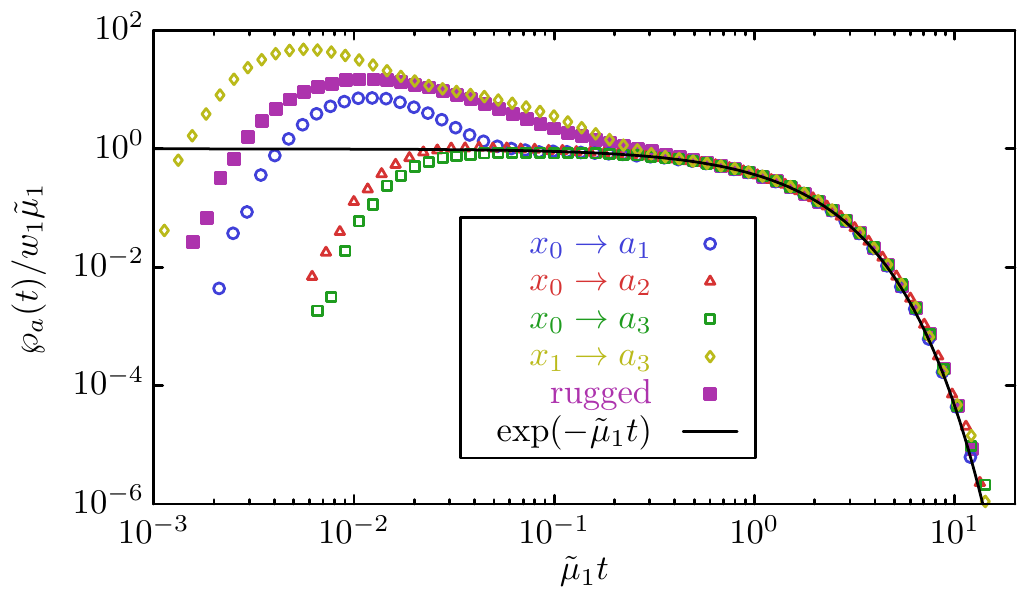}
\caption{Rescaled first passage time probability density, $\wp(t)/(w(x_0)\tilde{\mu}_1)$,
  obtained by rescaling the simulation results from Fig.~\ref{fg1}d (colored open symbols) and Fig.~\ref{fig_rugged}c (filled magenta rectrangles) by
  Eqs.~(\ref{weight}) and (\ref{principal}). The line denotes a
  unit exponential. }
\label{fg2}
\end{figure}

For single-particle problems the mean first passage time as well as
higher moments are typically
dominated by the long-time asymptotics of the first passage time distribution,
which we have also demonstrated in Fig.~\ref{fg_w}b for the triple-well potential.
The long-time limit is encoded in the principal first passage eigenvalue
$\mu_1$. As we demonstrate in \ref{Appendix1},
the principal eigenvalue $\mu_1$ can be obtained in a
simplified manner by formally setting $\bar\mu_k=0$ and $k^\ast=0$ in Eq.~(\ref{coefs}).
Moreover, a powerful approximation can be obtained by tuncating in
Eq.~(\ref{coefs}) all coefficients with $n>2$ (for a formal
justification see last paragraph of \ref{Appendix1}),
\begin{equation}
\mu_1\simeq\tilde{\mu}_1=\frac{\sigma_1(a)}{2\sigma_2(a)}\left[\sqrt{1+4\frac{P_{\mathrm{eq}}(a)\sigma_2(a)}{\sigma_1(a)^2}}-1\right],
\label{principal}
\end{equation}
where 
we introduced $\sigma_n(a)\equiv \sum_{l\ge 1}\langle a |\psi ^R_l\rangle\langle\psi
^L_l|a\rangle/\lambda_l^{n}$. 
Since Eq.~\eqref{principal} is derived from a Taylor expansion around
$s=0$ (see Eq.~\eqref{eq:A_F0}) it is expected to be quite accurate as
soon as the formal condition
$\mu_1\ll\lambda_1$ is met, which in turn translates self-consistently into $\tilde\mu_1\ll\lambda_1$. The relative error $\epsilon=|\mu_1-\tilde\mu_1|/\mu_1$ is expected to scale as $\epsilon\propto(\tilde\mu_1/\lambda_1)^2$.
For example, in the presence of at least one high barrier
and as long as $a$ is not the deepest point of $U(x)$ the condition $\lambda_1\gg
\mu_1$ is indeed satisfied (see, e.g. Fig. \ref{fg1}a,c).
Thus, rescaling $\wp_a(t)$
according to Eq.~(\ref{principal}), all curves must collapse for long times onto a unit exponential
$\wp_a(t=\theta/\tilde{\mu}_1)/(w(x_0)\tilde{\mu}_1)=\mathrm{e}^{-\theta}$, which is indeed fully
confirmed in Fig.~\ref{fg2}. The relative errors for the triple-well potential (see open colored symbols)
 are strictly bounded, $|\mu_1-\tilde{\mu}_1|/\mu_1<0.02$ for any $a$.

More generally, Eq.~(\ref{principal}) holds for
relaxation spectra obtained under a reflecting boundary at
$a,$ as well as for natural boundary conditions if there is no deeper minimum
beyond $a$.   If
furthermore $P_{\mathrm{eq}}(a)\to 0$ in Eq.~(\ref{principal})
(i.e. the case of 'rare-event' absorption), then $\tilde{\mu}_1\simeq P_{\mathrm{eq}}(a)/\sigma_1(a)$, where particularly
\begin{equation}
 \sigma_1(a)=\int_0^\infty [P(a,t|a)-P_{\mathrm{eq}}(a)]\dd t.
\end{equation}
Eq.~(\ref{principal}) generalizes the
`Poissonization' phenomenon observed in \cite{gode16,gode17}.  
We note that Eq.~(\ref{principal}) can also accurately describe the
long-time first passage asymptotics in rugged energy landscapes with
an arbitrary number of lower barriers (i.e. $<k_{\rm B}T$; see closed magenta rectangles in Fig.~\ref{fg2}). Further technical remarks including an
extension to discrete state systems can be found in \cite{hart18a_arxiv}.

\section{Conclusion}
\label{sec:conclusion}
This paper establishes rigorously the duality between relaxation and first-passage processes for ergodic reversible Markovian
dynamics. Based on the duality,
an intuitive explanation of first passage time
statistics in general rugged energy landscapes is provided. The full first passage time statistics are shown to be required for
explaining correctly the kinetics in the few-encounter limit -- particularly relevant cases
thereof are the triggering of diseases by protein misfolding and
related nucleation-limited phenomena. In addition, we obtained
accurate large deviation asymptotics dominating the mean first passage time, which emerge from a
time-scale separation in the relaxation process.
We show in \cite{hart18a_arxiv} that all concepts presented here can readily be extended to discrete state-space network dynamics, which, \textit{inter alia}
extends the duality between first passage and relaxation to higher dimensional networks.
Notably, they allowed us to determine, for the
first time, analytically the full first passage time statistics of the
Ornstein-Uhlenbeck process (see \cite{hart18a_arxiv}). Our work provides an exact unified framework for
studying the full statistics of first passage time under detailed balance conditions. Generalizations to irreversible
dynamics will be pursued in our future studies.

\ack

The financial support from the German Research Foundation (DFG) through the
\emph{Emmy Noether Program "GO 2762/1-1"} (to AG) is gratefully
acknowledged.

\appendix
\section{Proof of the duality}
\label{Appendix1}
In this appendix we sketch the proof of the duality, which allows us to determine
analytically the first passage time distribution from the
corresponding relaxation spectrum, i.e.,
the weights $\{w_k\}$ (not necessarily positive) and first passage rates $\{\mu_k\}$
that satisfy
\begin{equation}
 \tilde \wp_a (s|x_0)=\sum_{k>0}w_k(x_0)\frac{\mu_k}{\mu_k+s}
 \label{A1}
\end{equation}
directly from the relaxation spectrum $\{\lambda_k,\psiR_k\}$.
A detailed technical derivation including an extension to discrete state dynamics
can also be found in \cite{hart18a_arxiv}.
Since the first passage eigenvalues $\mu_k$ correspond to the poles
of $\tilde \wp_a (s|x_0)$ and the renewal theorem states
$\tilde \wp_a (s|x_0)=\tilde P(a,s|x_0)/\tilde P(a,s|a)$, our goal will be to find
the zeros of
\begin{equation}
 \tilde P(a,s|a)=\sum_{l\ge0}\frac{\langle a|\psiR_l\rangle\langle\psiL_l|a\rangle}{s+\lambda_l},
\end{equation}
where the reversibility of the Fokker-Planck operator
imposes $\langle a|\psiR_l\rangle\langle\psiL_l|a\rangle>0$
for all relevant relaxation modes $\langle a|\psiR_l\rangle\neq0$.

For the $k$th first passage rate $\mu_k$
we introduce the auxiliary functions
\begin{equation}
 F(k',s)\equiv  (s+\lambda_{k'})\tilde P(a,s|a),
 \label{eq:F}
\end{equation}
which for any $k'=k,k-1$ by design are strictly concave $\partial_s^2 F_{k'}<0$
within the interval $-\lambda_{k}<s< -\lambda_{k-1}$.
We choose
$k^\ast=k$ or $k^\ast=k-1$ such that $F(k^\ast(k),s)$ is restricted to be negative at $s=-\bar\mu_k=-(\lambda_k+\lambda_{k-1})/2$, i.e.,
\begin{equation}
k^\ast(k)=
\left\{
\begin{array}{ll}
k&\text{if $F_k(-\bar\mu_k)<0$},\\
k-1&\text{otherwise.}
\end{array}
\right.\label{eq:k*}
\end{equation}
Consequently, $F(k^\ast(k),s)$ is both negative and concave between
$s=-\bar\mu_k$ and $s=-\mu_k$, with $F(k^\ast(k),-\mu_k)=0$, which
implies that any Newton iteration starting from $\bar\mu_k$ will
strictly converge towards $\mu_k$.

The final step is to use an infinite Newton series -- an analytical
version of Newton's iteration --  in form of a series of almost triangular matrices \cite{gode16}.
First, we take the Taylor expansion of
\begin{equation}
 F(k^\ast(k),s)=\sum_{n=0}^\infty \frac{f_n(k)}{n!}(s+\bar\mu_k)^n,
 \label{taylor}
\end{equation}
where $f_n(k)=\partial_s^nF(k^\ast(k),s)|_{s=-\bar\mu_k}$, which are explicitly given in Eq.~(\ref{coefs}). Note that Eq.~(\ref{eq:k*}) implies $f_0(k)<0$. According to the interlacing theorem in Eq.~(\ref{interlacing}) the Taylor series in Eq.~(\ref{taylor})
converges on the full interval $-\lambda_{k}< s<-\lambda_{k-1}$
including the $s=-\mu_k$. Furthermore, the auxiliary function from Eq.~(\ref{eq:F}) and Eq.~(\ref{eq:k*})
guarantee the Newton series to converge to the true root $s=-\mu_k$ at which
$F(k^\ast(k),-\mu_k)=0$.
Hence, the $k$th first passage rate $\mu_k$ is exactly and explicitly given by the converging sum Eq.~(\ref{eigenv}).

Having determined the first passage eigenvalue $\mu_k$
the corresponding weight can simply be determined from
$\tilde \wp_a (s|x_0)=\tilde P(a,s|x_0)/\tilde P(a,s|a)$ and Eq.~(\ref{A1})
by
using the residue theorem that finally yields
\begin{equation}
 w_k(x_0)=\frac{\tilde P(a,-\mu_k|x_0)}{\mu_k\partial_s\tilde P(a,s|a))|_{s=-\mu_k}}.
 \label{Aweight}
\end{equation}
Eq.~(\ref{Aweight}) is equivalent to Eq.~(\ref{weight})
and
completes the proof since all first passage weights $w_k(x_0)$ and eigenvalues $\mu_k$
fully characterize the first passage time density $\wp_a(t|x_0)=\sum_{k>0}w_k(x_0)\mu_k\e^{-\mu_k t}$.

We finally provide some remarks on the principal eigenvalue $\mu_1$.
If only the principal eigenvalue is of interest the aforementioned discussion can be simplified
in the following way. First, we realize that there exist no relaxation
eigenvalue, which is smaller than $\lambda_0$. Therefore, the simple choice $k^\ast=0$
allows for a Taylor expansion of
$F(k^\ast{=}0,s)$ around $\bar\mu_{1}=0$ in Eq.~\eqref{taylor}
that converges on the full interval $-\lambda_1\le s\le +\lambda_1$,
which includes the lowest first passage eigenvalue $s=-\mu_1$.
Inserting $k^\ast=0$ and $\bar\mu_{1}=0$ in Eq.~\eqref{taylor}
and using Eq.~\eqref{coefs}
finally yields
\begin{equation}
  F(0,s)=P_{\rm eq}(a)+\sum_{n=1}^\infty(-1)^{n+1} \sigma_n(a) s^n,
  \label{eq:A_F0}
\end{equation}
where
$\sigma_n(a)\equiv \sum_{l\ge 1}\langle a |\psi ^R_l\rangle\langle\psi
^L_l|a\rangle/\lambda_l^{n}$, with $\mu_1$ being the negative zero of $F(0,s)$, i.e. $F(0,-\mu_1)=0$.
If we truncate the series \eqref{eq:A_F0} after $n=2$ we obtain a
simple parabolic equation with the solution
$\tilde\mu_1$ given in Eq.~\eqref{principal}.
Note that Eq.~\eqref{principal} equivalently follows from Eq.~(\ref{eigenv}) if one formally sets therein $f_n(1)=0$ for all $n> 2$ (see also \cite{hart18a_arxiv}).

\end{document}